# The Relationship between Precipitation and Aerosol: Evidence from Satellite Observation

Chongxing Fan, Maiqi Ding, Peipei Wu and Yaqi Fan

*School of Atmospheric Sciences, Nanjing University, China*

ABSTRACT

The interaction of aerosol-cloud-precipitation has an important impact on the global climate. The understanding of this issue is related to the uncertainty of climate change prediction. The traditional indirect effect of aerosols suggests that when the number of aerosols increases, it will act to suppress precipitation. However, recent studies on satellite observations have found that aerosols are positively correlated with precipitation, which is contrary to conventional views. This study attempts to use the A-Train satellite product to verify the correlation between aerosol and precipitation, and further reveals the possible physical mechanism of the positive relationship between aerosols and precipitation in satellite observations through the three-dimensional structure of clouds. The study found that the precipitation intensity is positively correlated with the aerosol optical depth, while the relationship between the cloud droplet concentration and the precipitation intensity is related to the liquid water path; and when the number of aerosols increases, the radar reflectance spectrum is widened, and the precipitation increases, while the cloud droplet concentration shows the opposite phenomenon. It can be concluded that the possible cause of positive correlation between aerosol and precipitation is the negative correlation between CDNC and AOD.

## 1. Introduction

Aerosol refers to a colloidal suspension system formed by dispersing solid or liquid small particles and suspended in a gaseous medium, also known as a gas dispersion system. The dispersed phase is a solid or liquid small particle having a size of 0.001 to 100 micrometers, and the dispersion medium is usually gas. The aerosol-cloud-precipitation interaction is a frontier issue in the field of aerosol and has an important impact on the climate. In recent years, with the increase of anthropogenic aerosol emissions and the increase of atmospheric pollution, the study of aerosol-cloud-precipitation interaction has become more and more important. However, there are still many uncertainties in the current understanding of this issue, which in part leads to uncertainty in our climate change projections (IPCC, 2013).

Aerosols can affect the weather through direct and indirect effects. The direct effect of aerosols means that aerosols directly affect the radiation balance of the earth-atmosphere system by absorbing and scattering solar radiation; while the indirect effect of aerosols refers to the influence of aerosols as cloud condensation nuclei (CCN) or ice nuclei (IN). Water content, cloud optical properties, cloud cover and cloud lifetime. Indirect effects can be divided into two categories: the first type of indirect effect, also known as the Twomey effect, refers to the increase in the concentration of aerosols, which causes an increase in the concentration of cloud particles and a decrease in the radius of cloud particles, thereby increasing the albedo of the cloud and affecting global radiation balance (Twomey, 1977); the second type of indirect effect, also known as cloud lifetime effect or Albrecht effect, means that the increase in aerosol concentration changes the microphysical properties of the cloud, and the reduction in the effective radius of the cloud droplets inhibits precipitation, The extension of the cloud's lifetime has an impact on the global radiation budget (Albrecht, 1989). Recent studies have also proposed a semi-direct effect of aerosols on clouds, such as black carbon or soot, which have a strong ability to absorb solar radiation and release heat radiation outward. The atmosphere and clouds make the cloud droplets evaporate, the amount of clouds is reduced, the cloud lifetime is shortened, and the average albedo of the cloud is reduced (IPCC, 2007).

The conventional cloud lifetime effect suggests that the reduction in cloud droplet size will slow the formation of rainfall. Among them, aerosol particles containing hygroscopic substances facilitate the condensation of water vapor on the surface, and such particles are called condensation nuclei. Precipitation that does not involve ice phase particles and cold cloud precipitation is called warm rain process. As a CCN (cloud condensation nuclei), aerosol affects the cloud droplet concentration and thus has a huge impact on the warm rain process. In the process of precipitation formation, it is very difficult for cloud droplets to grow into raindrops (radius greater than 100 microns) only by the condensation process. The presence of condensation nuclei makes it easy for water vapor to condense on its surface, forming some water droplets with diameters of several micrometers to tens of micrometers. In the formation of warm rain, the collision process plays a very important role. When the concentration of CCN increases, the concentration of cloud droplets will increase, and the radius of cloud droplets becomes smaller. The drop velocity and collision efficiency of cloud droplets decrease, which weakens the collision process and ultimately affects precipitation. Another effect of increased CCN concentration on the warm rain process is to narrow the cloud droplet spectrum. As the concentration of aerosol increases, the concentration of small droplets increases, resulting in a narrower scale spectrum of cloud droplets. The difference in cloud droplet size (decreasing speed) affects the collision process. The narrow cloud droplet spectrum will cause the cloud to collide and reduce the efficiency, and the weakening of the collision process will make the cloud droplet spectrum become narrower and finally suppress the warm rain process. It is inferred from the theory above that as the aerosol concentration increases, the precipitation will decrease.

In recent years, with the development of satellite remote sensing technology, it has become possible to study the interaction between clouds, precipitation and aerosols through satellites. The continuous observation of satellites can provide information on clouds, precipitation and aerosols worldwide. Through these data research and analysis, it is actually verified that the precipitation will decrease with the increase of aerosol. Huang Mengyu et al. (2005) used the Airborne Particle Measuring System (PMS) to investigate the effects of aerosols on clouds. The results show that there is a positive correlation between aerosol concentration and cloud droplet concentration. Ma Yue and Xue Huiwen (2012) used CloudSat and MODIS (Moderate Resolution Imaging Spectroradiometer) data to study the effect of aerosol on stratocumulus clouds. In the case of a certain liquid-water path in the cloud, the increase in aerosol can reduce the cloud droplet size. Zhao et al. (2006) analyzed MODIS data and station precipitation data, and pointed that the increase in aerosol caused the decrease in precipitation in eastern China.

However, recent satellite observation studies have found that aerosols and precipitation show a positive correlation under certain weather conditions. Wang et al. (2012) proposed the use of probability of precipitation (POP) as a function of aerosol to characterize this effect, and used the Aerosol Index (AI) to characterize the aerosol concentration, specifically defined as:

$$S_{POP} = -\frac{d\ln POP}{d\ln AI}$$

The previous work mainly showed the relationship between aerosol and precipitation through the observation of the MODIS satellite cloud of NASA in the United States and the rainfall observation of the TRMM satellite (Koren et al., 2014). Using this definition, the $S_{pop}$ calculated from the data collected by the satellite is close to 0, and even has a negative value, which indicates that the increase in aerosol concentration has little effect on precipitation, or even promotes, that is, aerosol and rainfall appear positive correlation, which is clearly contrary to classical theory.

Previous studies have also drawn relevant conclusions based on long-term observations. Menon, et al. (2012) studied the effects of black carbon aerosols on precipitation in southern China using a climate model. It was found that the heating effect of black carbon aerosols on the lower atmosphere increased the instability of the atmosphere and increased precipitation. Li et al. (2011) used long-term observations from the Great Plains of the United States to study the effects of aerosols on vertical motion and precipitation of clouds, and proposed an increase in aerosol concentration increases the precipitation in the case of higher liquid-water paths; while the precipitation is suppressed when the liquid water path is low. Other studies have pointed out that the effects of aerosols on clouds and precipitation are affected by meteorological conditions such as water vapor conditions, cloud top temperature and vertical velocity in the cloud but the relationship between the three is more complicated (Duan Wei and Mao Jietai, 2008; Guo Xueliang et al., 2013).

The analysis of the predecessors is based on observations, lack of analysis of the physical structure of the cloud, especially the detailed analysis of the three-dimensional structure, which makes the physical interpretation of the scientific phenomenon of positive correlation between aerosol and rainfall very insufficient. NASA's A-Train series of satellites simultaneously provide cloud and rainfall structures, especially fine-grained observations of three-dimensional structures, which provide observational data support for further understanding of this important scientific phenomenon. Our research program uses satellite observations of the three-dimensional structure of clouds and rainfall to further explore the scientific phenomenon of positive correlation between aerosol and precipitation, and expects to provide a mechanism explanation from the analysis of three-dimensional structures. The second section introduces the satellite data and processing methods used in this study. The third section shows the processing results and analysis conclusions. The fourth section summarizes the research conclusions and points out further research options.

## 2. Methods

The data used in this study is derived from the A-Train satellite of the National Aeronautics and Space Administration (NASA) Earth Observing System Program (EOS). The A-Train satellite formation includes Aqua satellites, CloudSat satellites, CALIPSO satellites, PARASOL satellites and Aura satellites. Each satellite has its own mission and is carefully teamed together for consistent observation and measurement results of higher quality and accuracy. The sensors used in the cloud research mission are mainly the Moderate Resolution Imaging Spectroradiometer (MODIS) on the Aqua satellite, the Cloud Profiling Radar (CPR) on the CloudSat, the Cloud Aerosol Lidar with Orthogonal Polarization on the CALIPSO satellite and Polarization and Directionality of the Earth's Reflectances on PARASOL (Jian Wei et al., 2008).

In verifying the relationship between aerosol and precipitation, and analyzing the three-dimensional structure of the cloud, we used CALIPSO's CAL_PM_L2 product to obtain aerosol information on the global cloud, including aerosol optical depth (AOD) and Ångström coefficient (AE). From these two quantities, the aerosol index (AI) can be derived as:

$$AI = AOD * AE$$

In addition, we use the MYD06_L2 and MYD08_D3 data on the MODIS satellite. The MYD06_L2 data is used to retrieve the liquid water path (LWP) and the cloud droplet number concentration(CDNC) at each grid point through the cloud optical thickness ($\tau$) and the effective radius of the cloud droplet ($r_{eff}$), ie:

$$LWP = \frac{2}{3}\rho_w \tau r_{eff}$$

$$CDNC = \alpha \tau^{0.5} r_{eff}^{-2.5}$$

Where $\alpha = 1.37 \times 10^{-5}\ m^{-1/2}$, $\rho_w$ is the density of liquid water (Brenguier et al., 2000; Quaas et al., 2006; Bennartz et al., 2007; Kubar et al., 2009; Suzuki et al., 2015).For the MYD08_D3 data, the global aerosol distribution information is extracted, and the physical quantity and calculation method are the same as the CALIPSO satellite.

Finally, we analyzed 2B-GEOPROF products and ECMWF-AUX products from the CloudSat satellite. The 2B-GEOPROF products provide radar reflectivity information for the global cloud. We found the value of the maximum radar reflectivity in each profile and used this as the cloud bottom radar reflectivity, using the Z-R relationship of Comstock et al. (2004) to retrieve the precipitation intensity R:

$$R = 2.01 * Z^{0.77}$$

The indicator POP for precipitation is defined as POP = 1 when the maximum radar reflectivity is greater than -15 dBz which means precipitation occurs, and POP = 0 when the maximum radar reflectance is less than -15dBz which means precipitation doesn't occur (Terai et al., 2012; Mann et al. , 2014; Terai et al., 2015); using the cloud mask value (CPR_Cloud_Mask) in the 2B-GEOPROF product, the clouded area can be screened for quality control. When the CPR_Cloud_Mask value is greater than 20, the cloud is judged in the sky. The greater the value, the greater the credibility, so the region with CPR_Cloud_Mask value less than 20 is excluded from this study. Only the role of warm clouds is considered in the study, so the warm clouds in the satellite data should be screened out. The ECMWF-AUX product provides temperature data at different heights, which allows the selection of warm clouds with a cloud top temperature greater than 273.15K.

## 3. Results

### 3.1. *Verification of positive correlation between aerosol and precipitation*

The main purpose of our study was to study the three-dimensional vertical structure of clouds and rainfall and its dependence on aerosols using satellite observations in the presence of a positive correlation between aerosols and precipitation. Negative correlation between aerosols and precipitation scenarios are used to reveal the possible underlying mechanisms of this phenomenon and better understand the interaction between aerosols and precipitation.

To study the changes in the three-dimensional vertical structure of the cloud when the aerosol is positively correlated with precipitation, it is necessary to verify the positive correlation between aerosol and precipitation. Koren et al. (2014) selected three regions of 9° × 9° in three different oceans whose latitude and longitude coordinates are:

Pacific Ocean：13° S ~ 22° S, 121° W ~130° W

Atlantic Ocean：17° S ~ 26° S, 19° W ~28° W

Indian Ocean：19° S ~ 28° S, 53° E ~62° E

By analyzing the TRMM satellite observation data from June to August 2007, a scatter plot of precipitation intensity and aerosol optical thickness is drawn. It can be seen that as the AOD increases, the precipitation intensity also increases, and they conclude that the aerosol is positively correlated with precipitation.(Fig. 1).

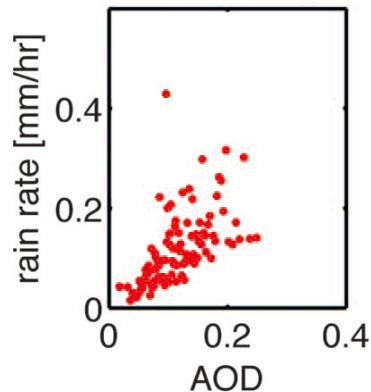

**Fig. 1.** Scatter plot of precipitation intensity and aerosol optical depth (Koren et al., 2014). AOD refers to aerosol optical depth, the more aerosol particles, the stronger the extinction effect and the larger the AOD; rain rate refers to the precipitation intensity, in millimeters per hour.

To verify this conclusion, we used observations from 2006 to 2011 to analyze these three regions through a multi-year average. In order to better illustrate the problem, we introduced other parameters that characterize aerosols, cloud droplets, and precipitation intensity. It can be seen from the figure that as the AOD increases, the precipitation intensity also increases, which is the same trend as in Koren et al. (2014). Note that the minimum value of the point is 1.5 (Fig. 2a), and the minimum value in the original paper is close to 0. The explanation for this is that the TRMM satellite is sensitive to moderate to heavy rain, while the CloudSat satellite is sensitive to light rain, so the value is large and There is a difference in distance.

The relationship between the precipitation intensity and the cloud droplet number concentration (Fig. 2b) and relationship between the precipitation intensity and AI (Fig. 2c) are also positive.

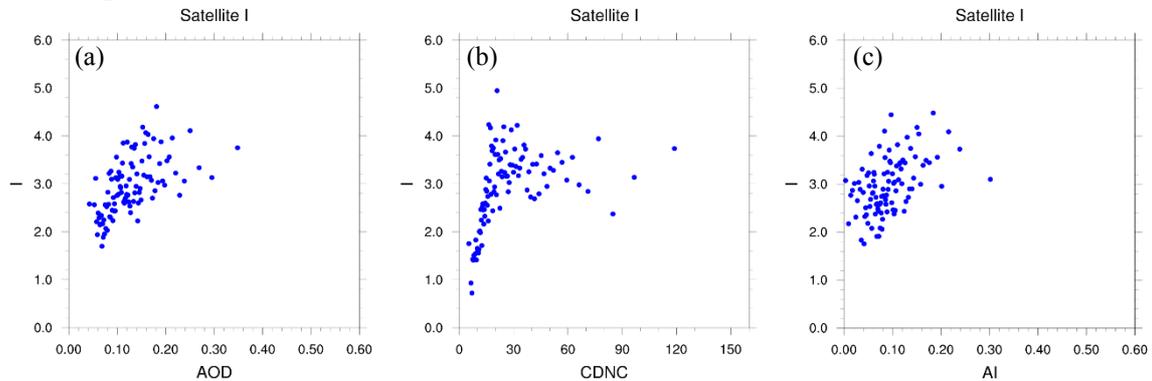

**Fig. 2.** Scatter plot of precipitation intensity and AOD (a), CDNC (b), AI (c), I refers to precipitation intensity (consistent with the physical meaning of the Rain Rate in the original literature, the unit is different: mm/h in the original literature, mm/24h in this paper); CDNC refers to cloud droplet number concentration; AI refers to aerosol index (AI is more precise than AOD in describing the aerosol number concentration)

In order to better understand this problem, we need to control meteorological conditions. Here we use LWP (Liquid Water Path, defined as the mass of liquid water per unit area of the air column). The data points were divided into 5 groups according to the LWP, and each group of data was divided into 10 parts, and the average precipitation intensity and average AOD/CDNC/AI of each piece were calculated. Precipitation intensity decreases with the increase of CDNC in the LWP large value region (Fig.3a), while in the LWP small value region Precipitation intensity increases slightly with the increase of CDNC. This is different from the trend shown in Figure 2. The reason is that all data points are averaged and does not distinguish between LWPs in Figure 2. Moreover, the larger the LWP, the more obvious the trend of negative correlation. The curve of I and AI is almost parallel to the X-axis (Fig. 3b), while the positive correlation between I and AOD in the full LWP range (Fig. 3c). This is consistent with the trend in Figure 2.

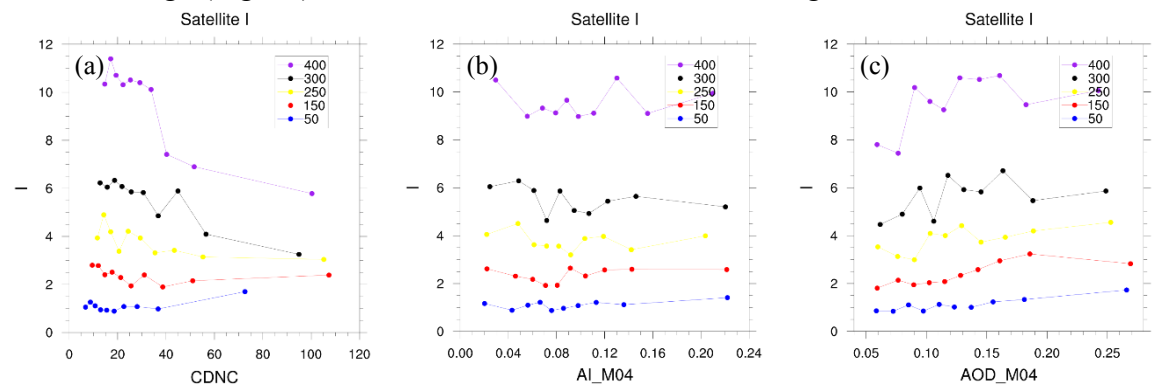

**Fig. 3.** Line charts of I and CDNC (a), AI (b), AOD (c). I refers to precipitation intensity. Each curve represents the relationship between I and CDNC/AI/AOD in a group of data in 5 groups. The number in the legend identifies the representative value of the set of LWPs.

The line chart in Figure 3 visually reveals the relationship, but we can also study by quantitative analysis. The concept of precipitation sensitivity is introduced here (Terai et al. 2015). The precipitation sensitivity used in this study is defined as:

$$S_x = -\frac{\partial \ln x}{\partial \ln CDNC}$$

Where x can be replaced by I, POP (probability of precipitation, POP = 1 means rain, POP = 0 means no rain) and R (precipitation intensity, defined as R = POP · I). The denominator CDNC can also be replaced by AOD and AI. When precipitation sensitivity is positive, precipitation and aerosol are negatively correlated; when precipitation sensitivity is negative, precipitation is positively correlated with aerosol.

According to the above definition, we draw the precipitation sensitivity maps of these three regions.

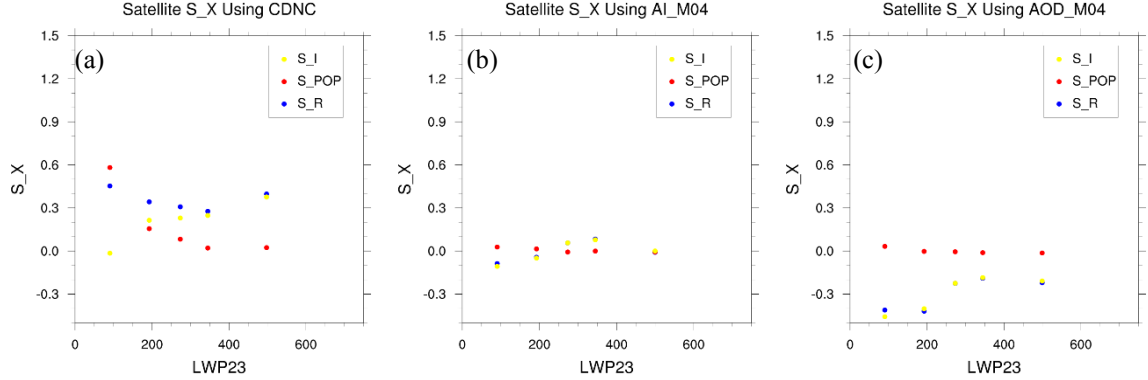

**Fig. 4.** Precipitation sensitivity scatter plot calculated with different denominators, (a) using CDNC, (b) using AI and (c) using AOD. In the legend, S_I, S_POP, and S_R represent $S_I$, $S_{POP}$ and $S_R$.

The equation $S_R = S_{POP} + S_I$ can be derived from the relationship R = POP · I. The sum of the subscripts. It can be derived from the relationship. It can be seen that the data points in Figure 4 are consistent with the formula. From the precipitation sensitivity plot of CDNC, except that the first point of S_I is close to zero, the other points all indicate positive values. This reveals that R, POP, and I both decrease with increasing CDNC, reflecting that an increase in the concentration of cloud droplets inhibits precipitation. This is consistent with conventional theories. From the precipitation sensitivity plot of AI, it can be seen that most of the points are near the value of 0, and the negative points are more. Combined with Figure 3, it can be found that the relationship between R, POP, I and AI is not obvious, and there is a slightly positive correlation. In the AOD precipitation sensitivity map, I and R have a large number of negative values, and the negative value is large. This shows that precipitation increases significantly with the increase of AOD. This is consistent with Figures 2 and 3.

We have validated the positive correlation between precipitation and aerosols from a number of perspectives (Koren et al. 2014), but there are still some questions need to be answered. How to explain the difference in precipitation sensitivity caused by using CDNC, AI and AOD as independent variables? We can establish the following relationship:

$$-\frac{\partial \ln x}{\partial \ln AOD} = -\frac{\partial \ln x}{\partial \ln CDNC} \cdot \frac{\partial \ln CDNC}{\partial \ln AOD}$$

The left side of the equation is the precipitation sensitivity with AOD as the denominator, and the first item on the right side is the precipitation sensitivity with CDNC as the denominator. The difference between the two depends on the second item to the right of the equation, the relationship between CDNC and AOD. It is known from traditional experience that aerosols can act as cloud condensation nuclei and promote cloud formation. So an increase in AOD will lead to an increase in CDNC. And we verified whether this theory is consistent with our satellite data, as shown in Figure 5.

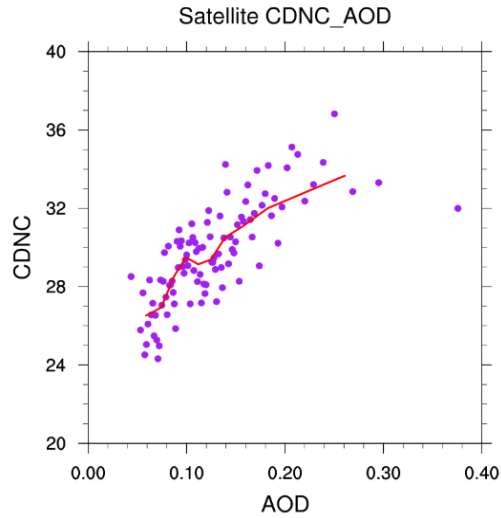

**Fig. 5.** Scatter plot of CDNC and AOD. Both the scatter and the fit curves in the figure indicate a positive correlation between CDNC and AOD.

From Figure 4, we know that the precipitation sensitivity of AOD as the denominator is negative, and the precipitation sensitivity of CDNC as the denominator is positive, which requires a negative correlation between CDNC and AOD, which is obviously contradictory. Similarly, according to the cloud water path, draw a line chart of the relationship between CDNC and AOD under different cloud water paths (Fig. 6).

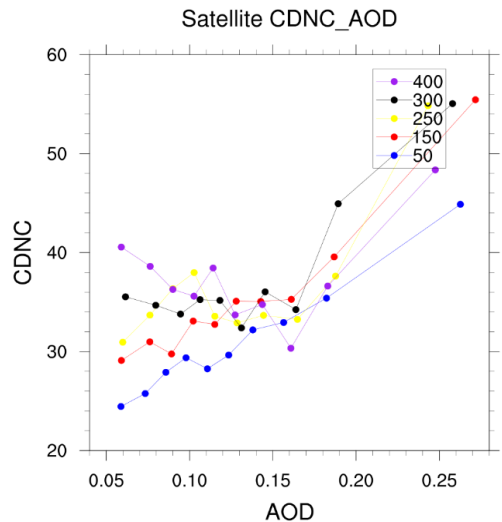

**Fig. 6.** Line chart of CDNC and AOD under different LWP conditions. When the cloud water path (LWP) is small (blue, red, yellow curve), CDNC increases with the increase of AOD. When the cloud water path (LWP) is large (black and purple curves), in addition to the two points when AOD is large, the previous trend is that CDNC has a negative correlation with AOD, which is consistent with the line graph in Figure 3.

Through different methods mentioned above, we can draw the following conclusions: 1. When the LWP is small, the precipitation intensity increases slowly with CDNC; when the cloud water path LWP is large, there is a clear negative correlation between the precipitation intensity and CDNC. 2. There is a positive correlation between I and AOD in the full LWP range. 3. The reasons for different results using AOD and using CDNC may be the negative correlation between CDNC and AOD in the high LWP area. Based on these conclusions, the positive correlation between aerosol and precipitation can be verified, which is consistent with the conclusion of Koren et al. (2014), and contradicts the conventional aerosol indirect effect theory.

## 3.2. *Analysis of the three-dimensional structure of the cloud*

In order to explore the physical mechanism of the above phenomenon, we tried to analyze the three-dimensional structure of the cloud in these three 9°×9° regions. We processed CloudSat Satellite's 2B-GEOPROF products and ECMWF-AUX products. 2B-GEOPROF products can provide radar albedo and CPR cloud cover indicators ( parameters determining the presence or absence of cloud) at different heights around the world, while ECMWF-AUX products can provide temperature information at different heights around the world. Through the temperature information and CPR cloud cover indicators, the warm cloud with cloud top temperature greater than 273.15K can be screened, and the frequency distribution histogram can be drawn by using radar albedo and altitude information. Then divide the area from 500m to 4000m with reflectivity factor ranging from -40dBz to 10dBz into 50 × 50 grids and average the results, which means the sum of all the frequency values on each level is 1.

It can be found that when AOD is smaller, the radar reflectance appears in the range at a relatively high frequency, and the spectrum is broaden. This reveals that when AOD is small, there is mainly clouds that generates small droplets, and does not form substantial precipitation. When AOD increases, the spectrum shifts to the right and precipitation increases (Suzuki et al., 2015). This is consistent with the conclusion 2 in the section 3.2. Similar conclusions can be drawn as well in different LWP conditions though not as obvious as Fig. 7a and Fig.7b due to the decrease in the number of data points.

Figure 8 and Figure 7 show the opposite situation. It can be found that when CDNC is larger, the radar reflectivity appears in the range at a relatively high frequency, and the spectrum is narrowed. That is to say, when the CDNC is increased, a small droplet cloud is mainly generated, resulting in less precipitation. This is consistent with the conventional physical mechanism explanation, because if the LWP is certain, the liquid water content is certain, then when the CDNC increases, each cloud droplet has less water content, so the droplet radius will reduce. However, this is inconsistent with the conclusions 1 and 3 in section 3.2, because Fig. 8a and Fig. 8b are both in small LWP value regions, which can not explain the positive relationship between I and CDNC when the LWP is small in Figure 3 or the phenomenon that CDNC and AOD in the small LWP value area are positively correlated in Figure 6. Because CDNC and AOD are difficult to retrieve, and the accuracy is affected by many factors, the reason for this strange phenomenon may be due to the retrieval error of CDNC and AOD, other aerosol parameters can be used for comprehensive analysis. In addition, due to the lack of large value of LWP, CDNC and AOD conditions, it is difficult to obtain enough observation data which increase the uncertainty.

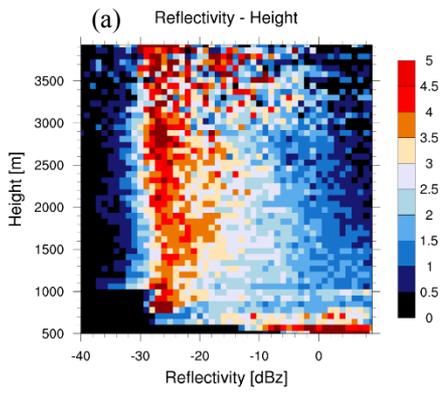
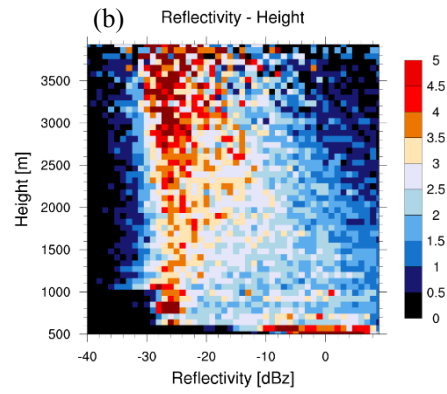

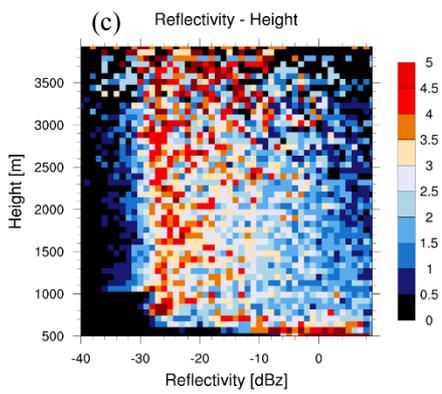
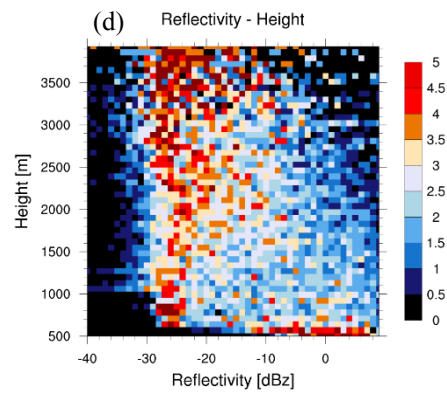

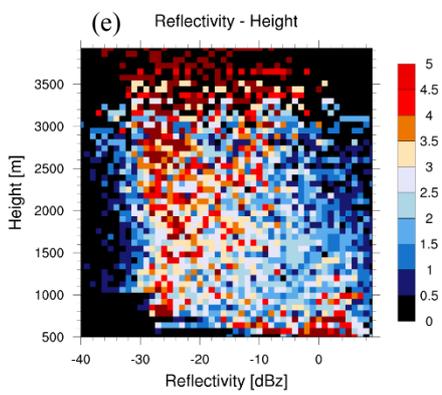
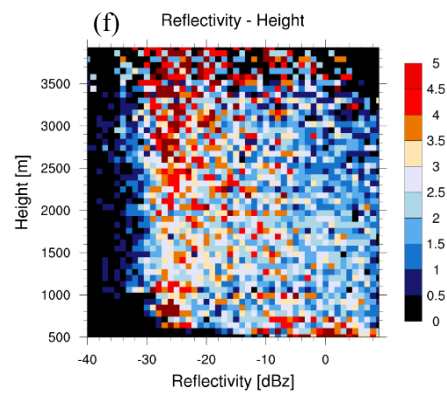

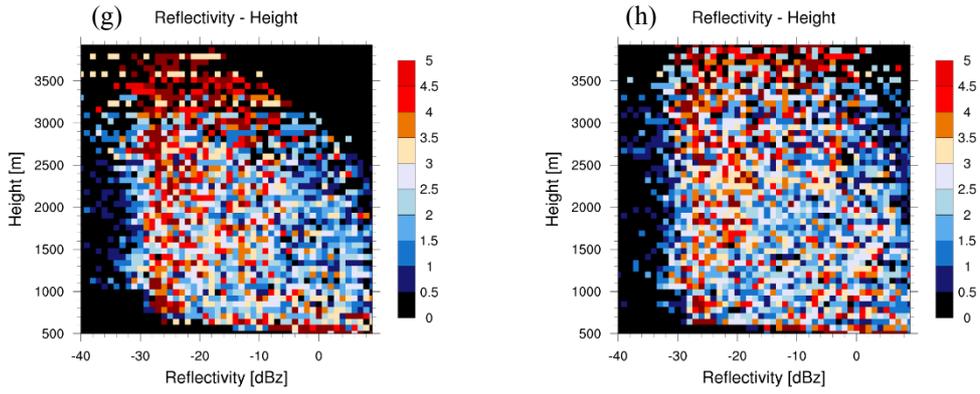

**Fig. 7.** Frequency distribution histograms with different range of LWP and AOD. In (a, c, e, g), AOD ranges from 0.00 to 0.10, and in (b, d, f, h), it ranges from 0.15 to 0.60. In (a, b), LWP ranges from 50 to 150, in (c, d), it ranges from 150 to 250, in (e, f), it ranges from 250 to 400, and in (g, h), it ranges from 400 to 700.

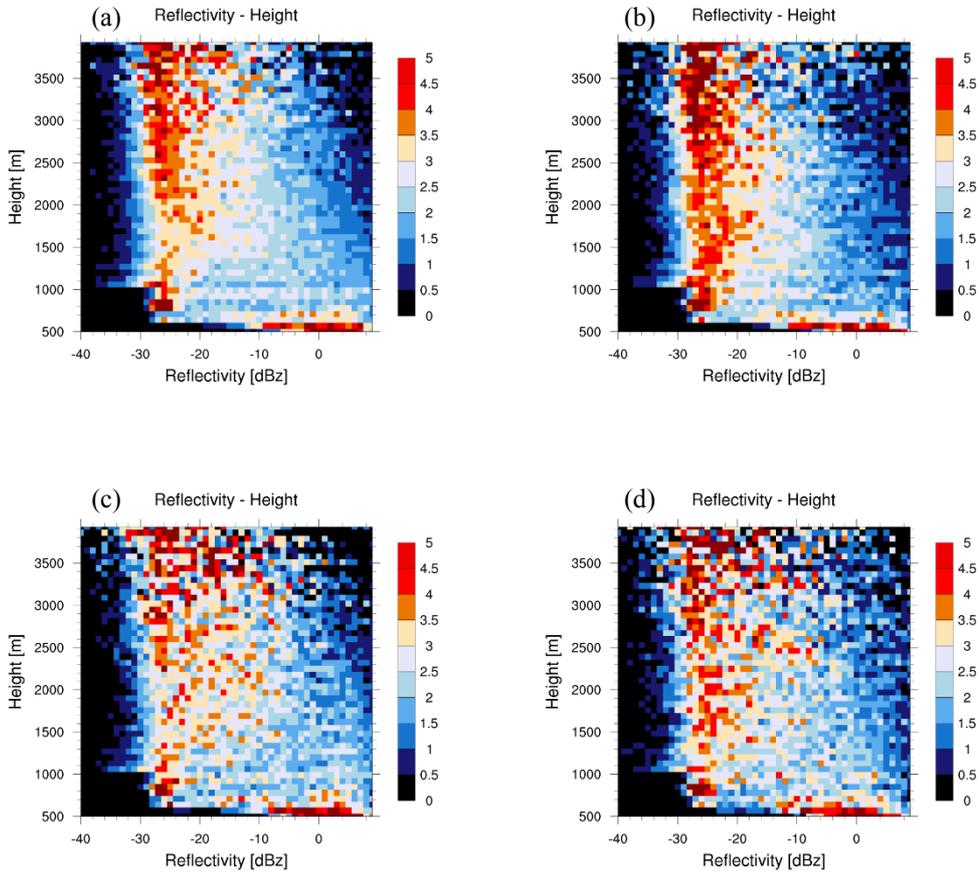

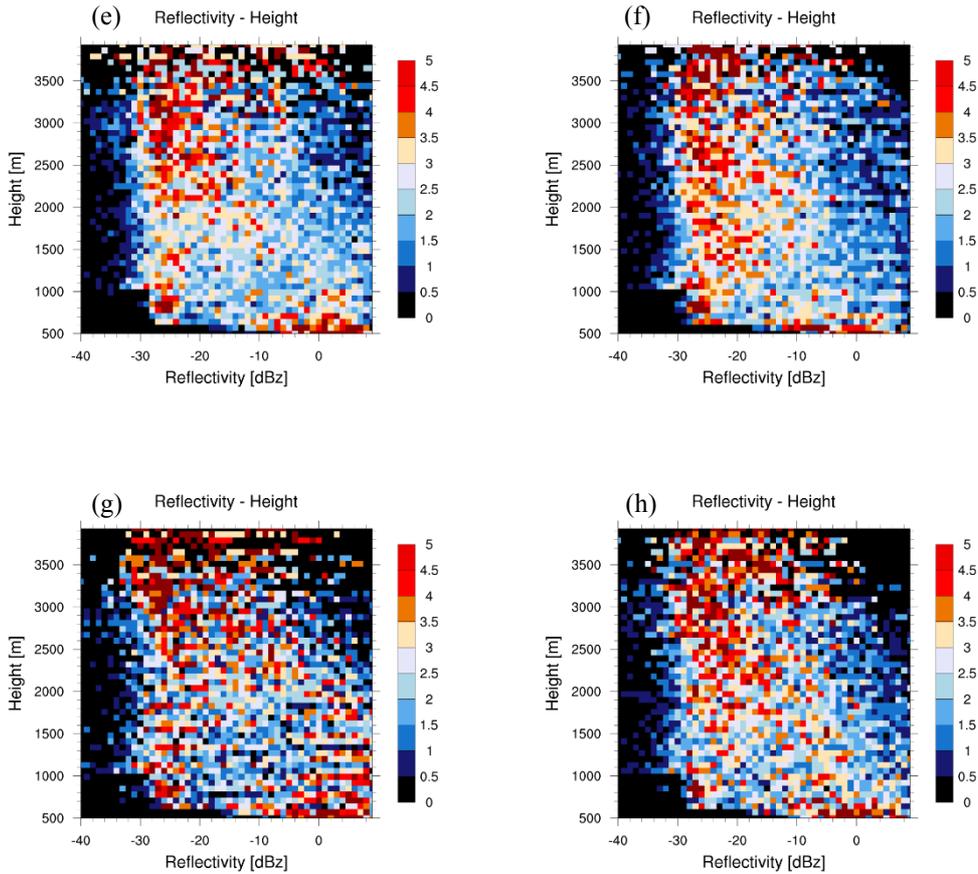

**Fig. 8.** Frequency distribution histograms with different range of LWP and CDNC. The grouping of LWP is consistent with Fig. 7. In (a, c, e, g), CDNC ranges from 0.0 to 20.0, in (b, d, f, h), it ranges from 30 to 200.

In general, from the three-dimensional structure of the cloud, it can be analyzed that when the aerosol optical depth increases, the reflectance spectrum is broadened, and the clouds that generate rainfall increase, resulting in the positive correlation between the previously determined precipitation intensity and AOD. When the cloud droplet number concentration increases, the reflectance spectrum narrows, the number of small droplet clouds increases, and the precipitation decreases, which is consistent with the conventional theory; the difference between the two may be caused by the retrieval error of CDNC and AOD and need to be further studied in the future.

## 4. Conclusion

The positive correlation between aerosol and precipitation is verified using the products of the A-Train series of satellites. The relationship between precipitation intensity (I) and aerosol optical depth (AOD) and cloud droplet number concentration (CDNC) were analyzed by selecting regions of the same size on three different oceans. In order to better explain the problem, we control the liquid water path (LWP) and analyze the correlation under different LWP conditions. These analyses verified the positive correlation between aerosol and precipitation intensity, and found the abnormal relationship between I and CDNC and AOD and CDNC. After this, we study the frequency distribution histogram of the radar albedo in the vertical direction through the analysis of the three-dimensional structure. It is found that the spectrum broadens and the precipitation increases when the number of aerosols increases, and the opposite conclusion occurs when the CDNC increases. Based on these conclusions, we infer that there are retrieval errors in the process of inverting CDNC and AOD physical quantities through satellite data, and it is suggested that caution should be exercised when using CDNC and AOD, but other aerosol parameters (such as AI) and clouds can be used. The parameters are comprehensively analyzed to compensate for the existing retrieval errors.

This study compared the reasons for the positive correlation between aerosol and precipitation intensity using the three-dimensional structure of the cloud compared to previous studies, and unlike using one-year TRMM satellite data, this study used A-Train series of satellite data from 2006 to 2011 which has been significantly improved in terms of data quality and accuracy, as well as statistical significance. However, due to the lack of large value of LWP, AOD and CDNC conditions, even if the multi-year comprehensive data is used, the image features become less obvious than the small-value regions. These can be further studied by using a combination of numerical simulations and satellite observation data.